# High-throughput assessment of vacancy formation and surface energies of materials using classical force-fields

Kamal Choudhary[1], Adam J. Biacchi[2], Supriyo Ghosh[1], Lucas Hale[1], Angela R. Hight Walker[2], Francesca Tavazza[1]

1 Materials Science and Engineering Division, National Institute of Standards and Technology, Gaithersburg, Maryland 20899, USA

2 Engineering Physics Division, National Institute of Standards and Technology, Gaithersburg, Maryland 20899, USA

## Abstract

In this work, we present an open access database for surface and vacancy-formation energies using classical force-fields (FFs). These quantities are essential in understanding diffusion behavior, nanoparticle formation and catalytic activities. FFs are often designed for a specific application, hence, this database allows the user to understand whether a FF is suitable for investigating particular defect and surface-related material properties. The FF results are compared to density functional theory and experimental data whenever applicable for validation. At present, we have 17,506 surface energies and 1,000 vacancy formation energies calculation in our database and the database is still growing. All the data generated, and the computational tools used, are shared publicly at the following websites https://www.ctcms.nist.gov/~knc6/periodic.html, https://jarvis.nist.gov and https://github.com/usnistgov/jarvis . Approximations used during the high-throughput calculations are clearly mentioned. Using some of the example cases, we show how our data can be used to directly compare different FFs for a material and to interpret experimental findings such as using Wulff construction for predicting equilibrium shape of nanoparticles. Similarly, the vacancy formation energies data can be useful in understanding diffusion related properties.

**Introduction**

Industrial materials often contain defects such as point defects (vacancies, interstitials, anti-sites), line defects (dislocations), planar defects (grain boundaries, stacking faults, twins) and bulk defects (pores, voids). As defects have direct influence on material properties like diffusion, catalytic activity, surface adsorption and free carrier concentration[1], a systematic investigation of defect behavior is essential for efficient materials design. Experimentally, it is challenging to develop a comprehensive and systematic database of these quantities compared to the ability to calculate these quantities using theoretical methods

Atomistic simulations based on classical[2] or quantum[3] mechanics principles are of immense importance in obtaining insight into defect-related material phenomena. Quantum mechanical calculations provide a much more accurate description of materials[4] than classical methods, but their applications are limited because their high computational cost restricts both the size and the time length of the simulations[5,6]. Classical mechanical simulations act as an aid to quantum mechanical tools to investigate large scale and defect related phenomenon such as the evolution or the effect of point, line and surface defects[7]. One of the biggest challenges in using classical simulations is that they are generally trained for specific applications and their applicability in other cases is unknown[6,8]. For example, if a force-field (FF) is trained for capturing elastic constants, their application to surface simulations are generally not easily predictable. Another major concern with classical force-fields is model verification and version control. Classical force-fields use a variety of functional forms, and each model has numerous empirically fit parameters, which could be implemented differently across different Molecular Dynamics (MD) or Monte Carlo (MC) computer-codes. Finally, reliable reference data is needed to validate the classical FF results, which can come either from experiments or higher-fidelity methods like density functional theory (DFT). As experimental data for all the materials and their prototypes are rarely available, DFT is in fact one of the most common resources to compare FF data.

Much work has been done towards assembling atomistic potential repositories and testing classical force-fields. The Interatomic Potential Repository (IPR) website[8] hosts force-field parameters verified by the developers. IPR classifies the potentials according to publication information, and lists all available implementation versions of each model. OpenKIM[9] uses an Application Program Interface (API) to design force-field models that combine code and parameters to make the implementations agnostic to the simulation software, and version control is handled at the model level. Both IPR and OpenKIM are also starting to add property calculations for their potentials. Other researchers and groups also host their own force-fields, with varying levels of version control such as https://sites.google.com/site/eampotentials/, http://www.ucl.ac.uk/klmc/Potentials/ and https://www.potfit.net/wiki/doku.php?id=potentials . However, presently all these databases mainly consist of computed data for metallic systems (mostly with Embedded Atom Method-EAM potentials). Therefore, the distribution and evaluation of advanced potentials and of potentials for non-metallic systems, such as Charge-Optimize Many Body (COMB)[10], Reaction Force-field (ReaxFF)[11] and so on, is still lacking. Additionally, the above-mentioned databases lack a link to reference values, from DFT for instance, to allow the user to directly evaluate the quality of the FF results. Providing an easy way for users to judge the applicability of a FF to their specific needs, and an array of FFs that goes beyond metallic compounds are key components of this work. Through the JARVIS-FF repository we provide a public database of simple and complex FFs. We also share a public framework to easily repeat and validate a calculation.

Previously[6] we evaluated energetics and mechanical properties of ideal materials with classical FFs. In the current work, we evaluate the vacancy-formation energies and surface properties using available classical force-fields. The vacancy-formation energy is a key component in predicting the activation barrier for self-diffusion[12], while the surface energies can be used to predict equilibrium shape of nanoparticles through Wulff-construction[4,13]. Poor prediction of surface energies will result in incorrect Wulf construction of equilibrium shape of nanoparticles for a system. Moreover, we provide the input and output files for all the calculations as well for the Large-scale Atomic/Molecular Massively Parallel Simulator (LAMMPS)[14] software, to further facilitate the user in running similar benchmarking tests. Please note that a commercial software is identified to specify procedures. Such identification does not imply recommendation by NIST. Our FF database is also linked to DFT databases such as Materials-Project (MP)[4] and JARVIS-DFT[15] to help the user assess a FF applicability to a specific property calculation. Materials-Project and JARVIS-DFT consists of nearly 70,000 and 30,000 materials, respectively, evaluated with density functional theory and both are continuously expanding. Some of the calculated properties in these databases are heat of formation, elastic constants, defect formation energies, surface energies and phonons, which can be considered as reference data for FF results. Both JARVIS-FF and JARVIS-DFT are part of Joint Automated Repository for Various Integrated Simulations (JARVIS) at NIST. Both projects are linked through unique identifiers for calculations (discussed later).

The paper is organized as follows: first the methods for developing this project are discussed, followed by results and discussion. Example cases are discussed to elaborate the applications of our database. Comparisons to experimental results are also provided for several cases. Finally, conclusions and future work are discussed.

**Methods:**

The database creation consisted of several steps. A flow-chart of the processes involved is shown in Fig.1.

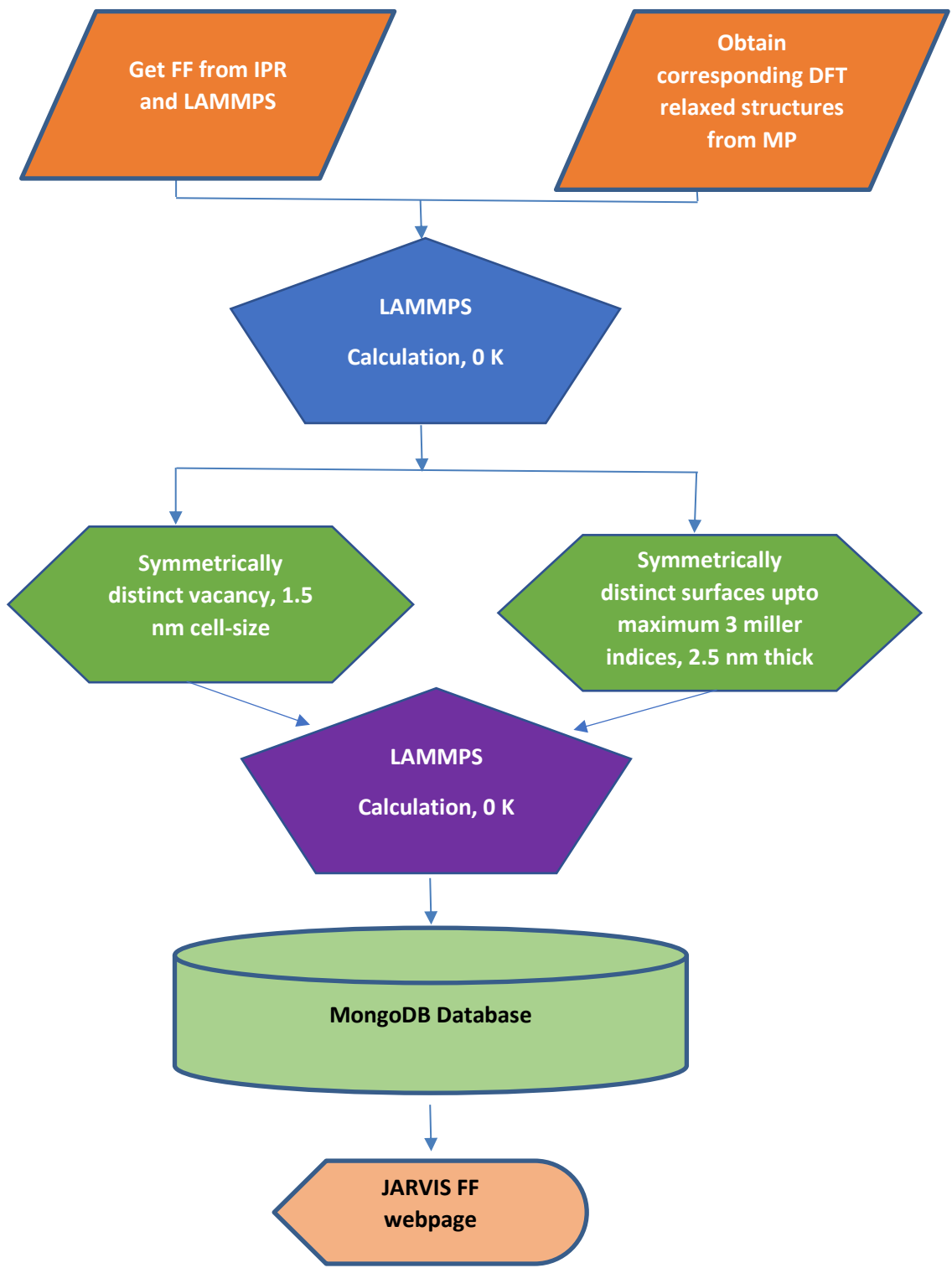

*Fig. 1 Flow-chart showing the processes involved in creating the JARVIS-FF database.*

Crystal structures were mainly obtained from the Materials-Project (MP) [4] database. The force-fields/interatomic potentials were obtained from LAMMPS[14] and NIST's interatomic potential repository[8]. For a material or system of interest, such as Ni-Al, the corresponding crystal structures were obtained from MP. The crystal structures were then converted into LAMMPS format using python tools, which are publicly available at our JARVIS github page (https://github.com/usnistgov/jarvis). The script uses pymatgen[16], ase[17] and pydii[18] codes. It is important to note that the MP crystal structures are generally available in DFT code format, which need to be converted to LAMMPS format prior to running the calculation using our scripts available at github. The LAMMPS input files (with simulation controlling parameters) corresponding to the crystal structure file were generated and subjected to a computer queuing system. In our structure minimization calculations, we used $10^{-10}$ eVÅ$^{-1}$ for force convergence and 10000 maximum iterations. The geometric structure is minimized by expanding and contracting simulation box with fix box/relax command and adjusting atoms until they reach the force convergence criterion. These are generalized computational set-up parameters. After the minimization, the crystal structure is stored in LAMMPS data-format and JSON format. Using this JSON file, unique Wyckoff positions were identified and deleted to represent vacancy-structures. The multiplicity of the Wyckoff positions is also recorded. After the defect structure generation, the LAMMPS energy minimization is carried out. In a subsequent run, we calculate the chemical potential of the defect element using the specific force-field. The structure for the most stable prototype of the element was obtained from MP. The data for the vacancy structure, chemical potential of element and perfect structure energy were used to calculate the defect formation energies. The most stable prototype for chemical potential calculation was determined using the energy above convex hull data from DFT. The defect structures were required to be at least 1.5 nm long in the x, y and z directions to avoid self-interactions with the periodic images of the simulation cell. Similar to the defect structures, distinct surfaces were created up to 3 Miller indices with the relaxed structure stored in the JSON file. A generic code for generating defect and surface structures is given at our github page. We enforce the surfaces to be at least 2.5 nm thick and with 2.5 nm vacuum in the simulation box. The 2.5 nm vacuum is used to ensure no self-interaction and the thickness is used to mimic actual experimental surface. Using the energies of perfect bulk and surface structures, surface energies for a specific plane are calculated. We should point out that only unreconstructed surfaces without any surface-segregation effects are computed, as our high-throughput approach does not allow for taking into account specific, element dependent reconstructions yet.

Although specific LAMMPS simulation set-up parameters, and structure creation parameters are required here, our scripts are completely flexible to utilize any user input. All the data generated were stored in a MongoDB format in www.jarvis.nist.gov and in supplementary information. This database was also used to make the user-friendly webpages on www.ctcms.nist.gov/~knc6/periodic.html. The webpage was created based on Javascript and the crystal structure visualization was supported with the JMOL software package.

For experimental validation, gold, platinum and silver nanoparticles were experimentally synthesized using a heat-up route employing metal salt precursors and a solvent that also acted as the reducing agent[19]. All syntheses were performed in a 25 mL three-neck flask fitted with a condenser, a septum, and a glass-coated thermocouple. The reaction temperature was directed using a digital controller connected to the thermocouple (J-KEM Scientific) and a 25 mL heating mantle (Glas-Col). Please note that a commercial software is identified to specify procedures. Such identification does not imply recommendation by NIST. The solution was stirred vigorously with a Teflon stir bar and continuously sparged with flowing argon. For Pd nanoparticles, $Na_2PdCl_4$ (0.1 mmol, > 99.9 %, Sigma-Aldrich) and poly(vinylpyrrolidone) (PVP, 2 mmol by repeating unit, MW = 40,0000, TCI) were dissolved in 10 mL of ethylene glycol (EG, >99 %, JT Baker) and heated to 95 °C for 1.5 hours. Pt nanoparticles were synthesized by dissolving $H_2PtCl_6 \cdot 6H_2O$ (0.1 mmol, 99.9 %, Alfa Aesar) and PVP (2 mmol) in 10 mL of EG, then heated to 150 °C and held for 1.5 hours. For Au nanoparticles, $HAuCl_4 \cdot 3H_2O$ (0.05 g, 99.99 %, Alfa Aesar) was dissolved in 5 mL oleylamine (70 %, Sigma-Aldrich) and 5 mL octadecene (90%, Sigma-Aldrich), then heated at 105 °C for 45 min. Pd and Pt nanoparticles were precipitated by centrifugation in an excess of hexanes and then re-dispersed in ethanol. Au nanoparticles were precipitated likewise, using an excess of ethanol before redispersion in toluene. Transmission electron microscopy (TEM) images were collected using a Phillips EM-400 operating at an accelerating voltage of 120 kV and high-resolution TEM (HRTEM) images were obtained from a FEI Titan 80-300 operating at 300 kV. Samples were prepared by casting one drop of dilute nanoparticle solution onto 300-mesh Formvar and carbon-coated copper grids (Ted Pella).

**Results and discussion:**

All data computed here can be found at our database website ( https://www.ctcms.nist.gov/~knc6/periodic.html). Our database contains 17,506 surface-energy and 1,000 vacancy formation energies calculation for 107 force-fields and 1215 materials and the database is still growing. Our computational framework is designed to automatically update on the website with the FF calculation results when a new force-field is added to our database. A snapshot of the database web-page is given in Fig. 2. This is similar to our webpage for elastic constants calculations that were presented in our previous work[6]. A user can enter an element or element combination in the interactive periodic table, such as Al, or Ni-Al by clicking on Al or Ni and then Al in the periodic table. Now, as the user clicks on 'search' button, a result table appears at the bottom of the periodic table that is a concise summary of the material properties and contains links to the detailed webpages. The first column in the table contains the JARVIS-ID, which is hyperlinked to a detailed webpage containing the computed properties for that specific material and force field. The second column shows the chemical formula, the third and fourth columns display the space-group and the force-field, respectively. Bulk and shear modulus ($B_v$, $G_v$) are given in the next two columns. Minimum surface and vacancy energy are shown next, whenever available. Finally, MPID: materials project (MP) is hyperlinked.

Clicking on JARVIS-ID (for example: JLMP-1243) navigates to a detailed webpage containing details of the material, force-field, a structural visualization tool based on JMOL, elastic tensor data, vacancy formation energy data, surface energy data and Wulff construction plot. The 'Download input files' button gives the user access to the input files (provided in zipped format) to compute elastic properties for the ideal system. Detailed files to create the surfaces and defects are not given for each material, but scripts to create them are provided on our github page.

All the major approximations used in the calculations are listed in each detailed webpage. Vacancy formation energies with symmetrically distinct sites are shown on the webpage along with the site-multiplicity. This data can be compared to more sophisticated computational results such as DFT to evaluate the validity of the force-field for vacancy-related calculations such as diffusion barrier calculations. Similarly, the unreconstructed surface energies along with the Miller indices are given in increasing order. Generally, the order of surface energy values should be comparable to DFT results to examine surface-related phenomena such as ion-beam sputtering and catalysis. A Wulff-construction plot for the material using the particular potential is also given. At present, we have Embedded Atom Method (EAM), Modified Embedded Atom Method (MEAM), Tersoff, Reactive Empirical Bond Order (REBO), Adaptive Reactive Empirical Bond Order (AIREBO), Stillinger-Weber (SW), Streitz, Charge-Optimized Many Body (COMB), Reaction Force-field (ReaxFF) and Spectral Neighbor Analysis Potential (SNAP) force-fields[14]. It is to be noted that our database is still in the development phase, and some of the Wulff-construction data are still missing. As calculations are completed, the webpages will be automatically updated.

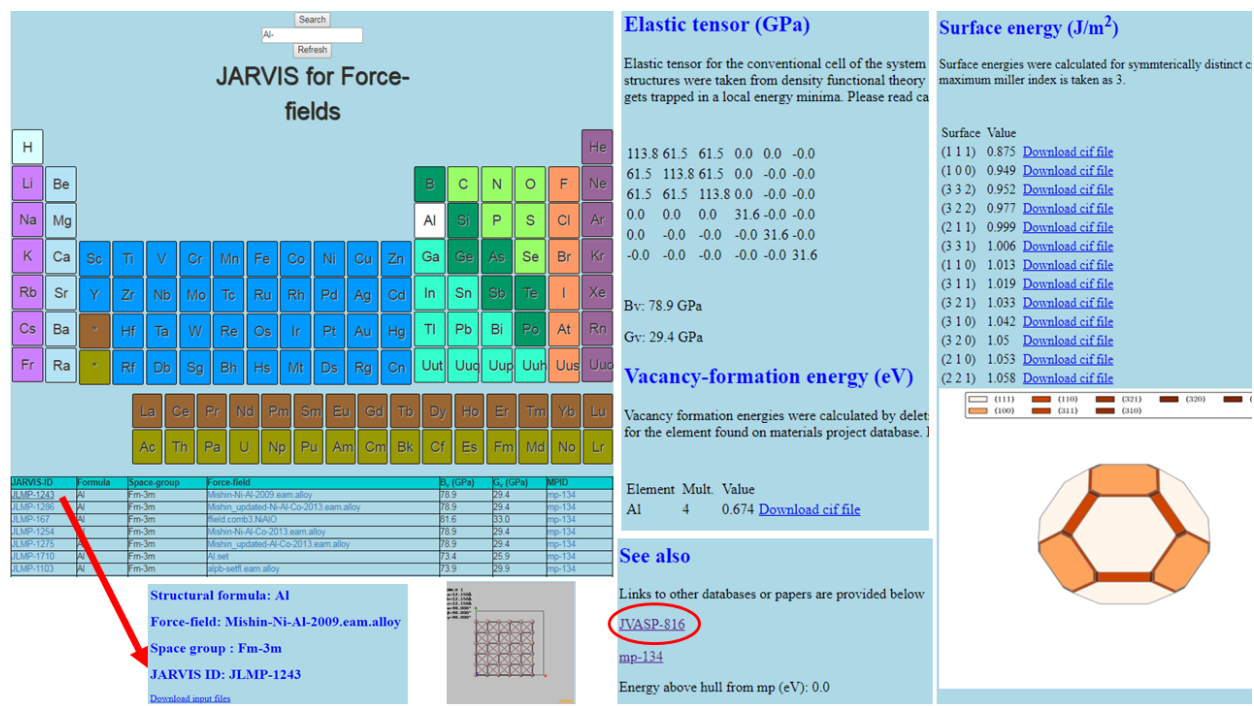

*Fig. 2 Snapshot of the database showing the interactive periodic table and an example output web-table for 'Al' with JARVIS-FF identifier JLMP-1243 are shown. There are similar webpages for all the available material and available FF combinations. The JARVIS-DFT link is highlighted by the red ellipse in the 'See also' section.*

As mentioned previously, it is important to asses a force-field by comparing it to DFT or experimental data before carrying out a specific MD simulation. DFT data are more readily and digitally available than experimental ones, hence DFT data are used for comparison here. As an example, we compare some of the vacancy and surface energy data for all single elements (with available FF) in Figs. 3a and 3b respectively. The DFT data were obtained from previous work by Medasani et al[20]. In general, force-fields giving opposite sign for vacancy formation energies with respect to DFT results should not be used to compute defect-related properties. It should be noted that generally very few symmetrically distinct sites are available for vacancy formation compared to free surface creations, hence, the data in Fig. 3a (vacancy formation energies) is much less than Fig. 3b (elemental surface energies). Vacancy formation energies in semiconductors and insulators may be particularly difficult to correctly predict with FFs due to the existence of charge states[21-23]. The charge induced defect properties are captured well using DFT but can't be reproduced classically. Additionally, a user comparing FF results with DFT should also be careful about finite temperature effects and different settings or functionals in DFT calculations. With respect to temperature, as DFT data are computed at T=0K, comparing finite temperature FF results to them directly may be misleading. Furthermore, most FFs are fitted on a combination of experimental data (usually at room temperature) and DFT values (T=0K). This can lead to some inconsistency inside the potential, of which the user should be aware of.

As shown in Fig. 3a, the vacancy formation energies can be higher or lower compared to DFT data, but Fig. 3b shows that force-field surface energy data are generally underestimated compared to DFT. This typically implies that a surface can be more easily formed in MD simulations than in DFT. Commonly, energies related to surfaces with Miller indices up to 1 are included in the force-field training data, so good agreement with DFT should be expected. We computed energies for unreconstructed surfaces up to Miller index 3, which may explain the large differences with the DFT data in Fig. 3b.

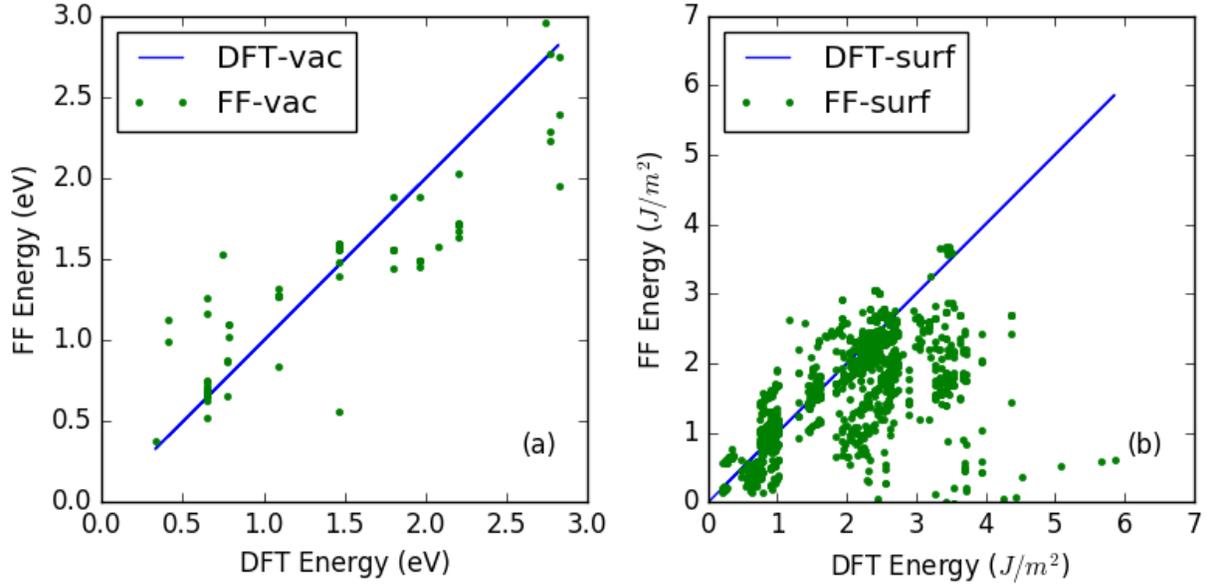

*Fig. 3 Vacancy and surface energies (up to a maximum Miller index of 3) for single elements obtained with all the FF available for such materials are compared to available DFT data.*

As an example, in Fig. 4 we analyze the specific case of Al out of all the single elements described in Fig. 3. Vacancy formation energies and surface energies for the (111) surface of face-centered cubic (FCC) Al obtained with various force-fields available in our database are compared. We find that FFs such as NiAl02_eam. alloy[24] and Farkas Nb-Ti-Al_1996.eam.alloy[25] potentials overestimate the vacancy formation energies for Al, while NiAlH_jea.eam.alloy[26] underestimates the vacancy formation energies. This was not completely unexpected, as many FFs are not designed to predict accurate vacancy formation energies. Defect energetics play a pivotal role in determining the diffusion behavior of a material, hence a FF with accurate defect formation prediction is generally needed for describing materials phenomenon. A FF with negative vacancy formation energies implies that the FF could show unphysical behavior compared to DFT. To expand the comparison from just elemental to binary systems, we compare the FF vacancy formation energies of few binary systems to DFT in Table 1. Based on this small dataset, we find that the vacancy formation energies are generally overestimated with FFs. Much deviation is observed in surface energies compared to vacancy formation energies. FFs such as Al-Fe.eam.fs and Al-Mg.eam.fs[27] can underestimate the surface energies up to 30%. Although we show example of Al vacancy and (111) surface energy in Fig. 4, other possible element combinations can be analyzed with our database.

Table 1. Comparison of vacancy formation energy for binaries between density functional theory (DFT) and force-field (FF). The FF identifiers are specified as JLMP-# for the detailed webpages of the calculations.

| NiAl (Pm-3m) | DFT (eV) | FF (eV) | FF (eV) |
| --- | --- | --- | --- |
| $V_{Al}$ | 2.14[a] | 2.03, JLMP-1245 | 3.06, JLMP-1330 |
| $V_{Ni}$ | 2.18[a] | 2.85, JLMP-1245 | 2.90, JLMP-1330 |
| **Ni₃Al (Pm-3m)** | | | |
| $V_{Ni}$ | 1.95[b] | 2.86, JLMP-1244 | 2.74, JLMP-1352 |
| $V_{Al}$ | 2.81[b] | 3.36, JLMP-1244 | 3.41, JLMP-1352 |
| **TiAl (P4/mmm)** | | | |
| $V_{Ti}$ | 1.54[c] | 3.2, JLMP-1171 | |
| $V_{Al}$ | 0.55[c] | 2.0, JLMP-1171 | |

a) Ref[28]
b) Ref[29]
c) Ref[30]

Although we discuss the specific example of Al for a single vacancy and one surface energy (Fig. 4) only, many other possible combinations are provided in our database, and a user can/should perform such comparisons before carrying out an actual classical force-field calculation.

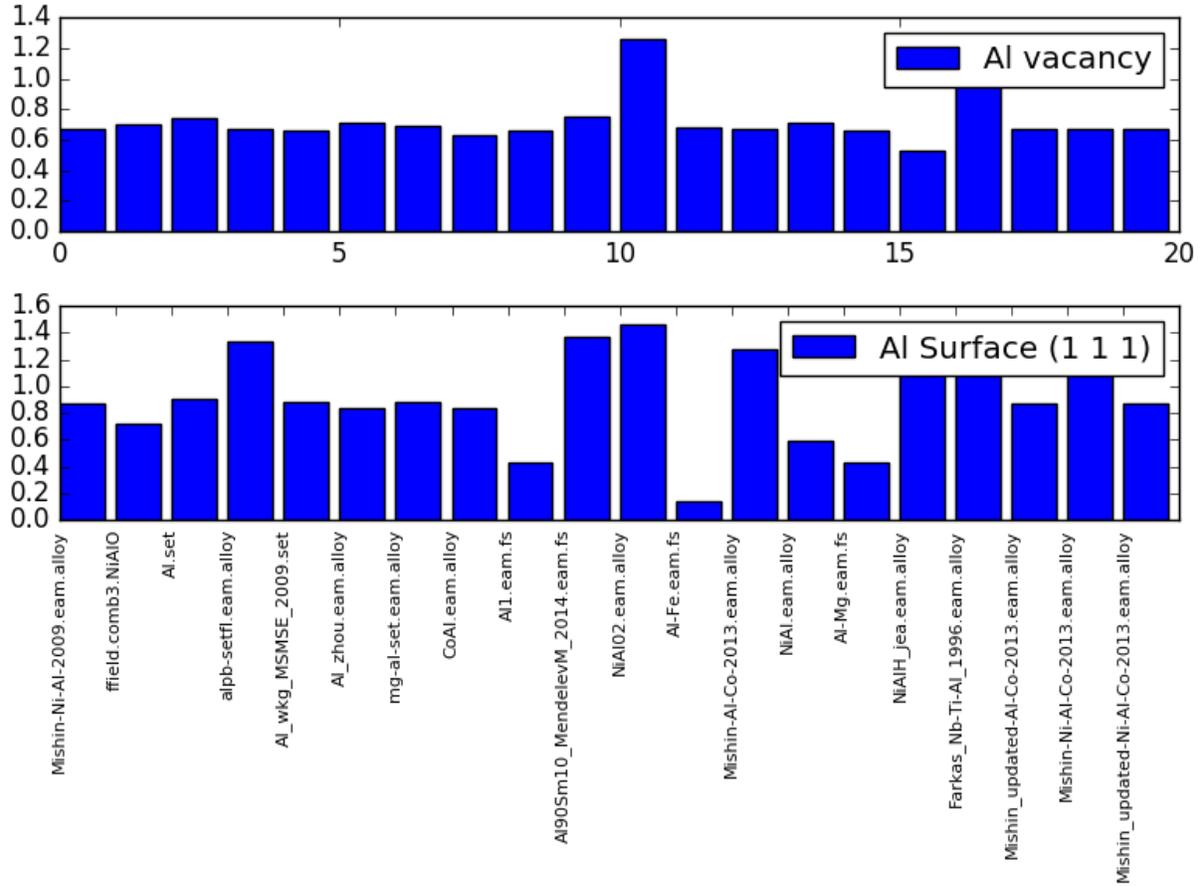

*Fig. 4 Comparison of Al vacancy-formation energies (eV) and surface energies for (111) (Jm$^{-2}$) plane using various FFs[24-27,31-41].*

Examples of calculated single-element equilibrium surface structures are displayed in Fig. 5, indicating that energy is minimized for FCC metals such as Pd, Pt and Au through the formation of a cuboctahedral-geometry crystallite bound by a mixture of (111) and (100) facets. This model was first compared with DFT data obtained from ref. [4]. The Pd and Pt Wulff plot obtained for classical calculations agree with that from DFT plot, but Au predictions are very different. DFT predicted a non-cuboctoctahedral geometry for Au, which is counter-intuitive. The Au DFT results clearly show that DFT Wulff data are also not very reliable. This may be due to selection of specific functional or DFT parameters. In general, DFT has limitation in reproducing certain physical quantities such as surfaces energies[42,43] (11% underestimation of the surface energy by PBE), defect formation energies[20] (mean absolute percentage deviation 0.43) and mechanical properties[44,45] (bulk modula could be off as much as ±15% with respect to experimental values). In this respect, the FF data can be more useful if fitted to experimental data. Additionally, our computational framework allows us to study the temperature and size effects of the surfaces on Wulff plot, which is very difficult to investigate using DFT. The temperature dependent evaluation of FFs will be carried out later.

Next, we compare our findings with actual experimental results obtained from reducing complexes of FCC metals in the presence of colloidal-stabilizing agents to form nanoparticles using a standard heat-up method[46]. As shown in Fig. 6a-c, TEM analysis indicates Pd, Pt, and Au all selectively form cuboctahedra in high yield, which appear roughly spherical at moderate magnifications. Moreover, HRTEM imaging (Fig. 6d-f) confirms that these particles are bound by a mixture of (111) (white outline) and (100) (red outline) surfaces. Fast Fourier transforms of these regions verified the expected hexagonal in-plane packing geometry of the (111) facet and the 90° arrangement of atoms relative to each other that is characteristic of FCC (100). Importantly, these nanoparticles crystallized under mild reducing conditions, which promote the formation of thermodynamically favorable morphologies. Although anionic species can potentially influence the energetic favorability of certain metallic facets by means of surface coordination, none of the potential ligands present during these syntheses (including ethylene glycol, oleylamine, and Cl$^-$) have a strong adsorption affinity to these noble metal surfaces according to hard-soft acid-base theory[19]. It is also noteworthy that similar structural motifs have been observed in alloys of FCC metals as well[47].

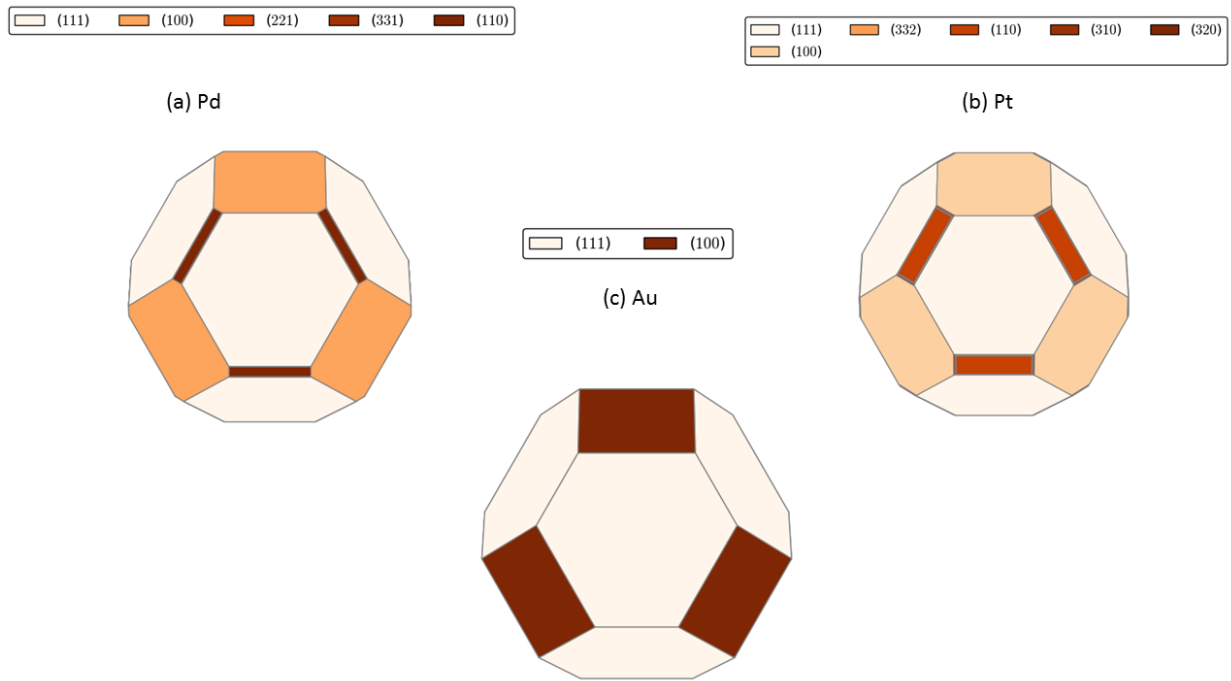

*Fig. 5 Equlibrium shapes of Pd, Pt and Au nanoparticles using classical force-fields[33,48], indicating the lowest-energy surface configuration results are bound by (111) and (100) facets. Example webpages for the above results can be found at https://www.ctcms.nist.gov/~knc6/jsmol/JLMP-1753.html for Pd , https://www.ctcms.nist.gov/~knc6/jsmol/JLMP-1754.html for Pt and https://www.ctcms.nist.gov/~knc6/jsmol/JLMP-1714.html for Au.*

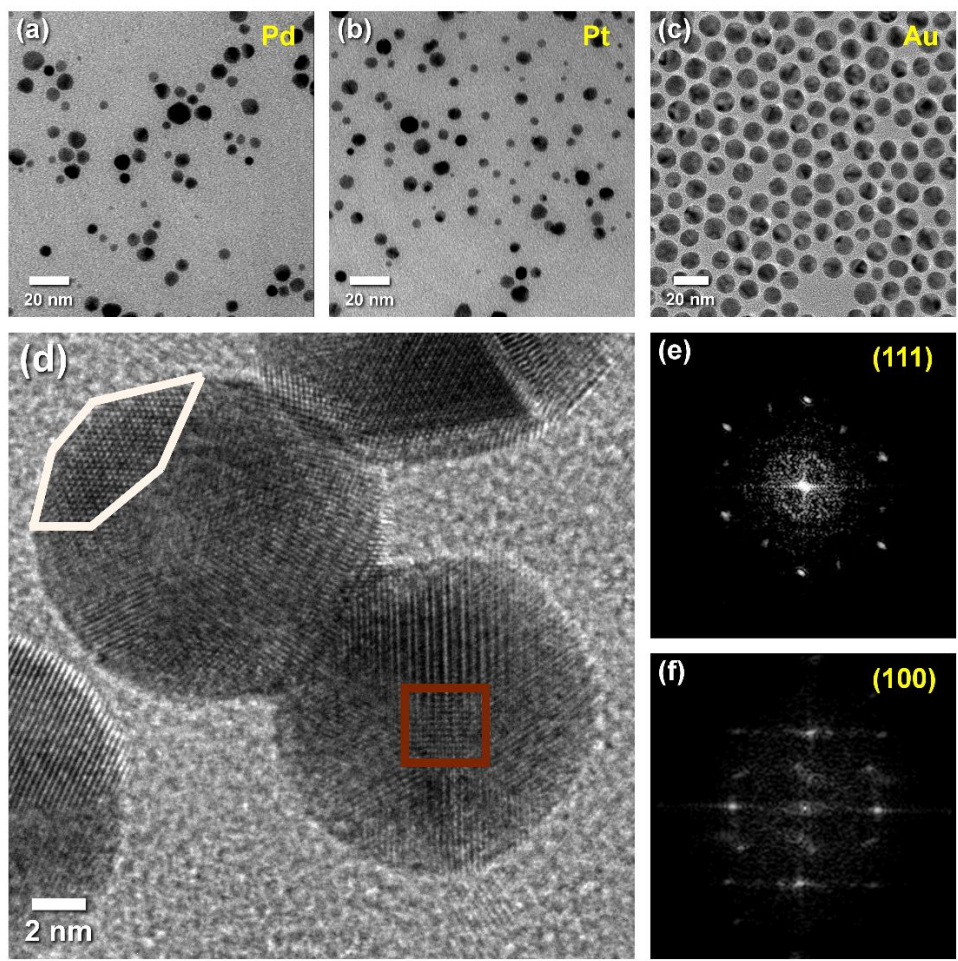

*Fig. 6 TEM images of (a) Pd, (b) Pt, (c) Au cuboctahedral nanoparticles synthesized by reducing corresponding metal salts in the presence of weakly binding colloidal stabilizers. (d) HRTEM analysis of the several Au nanoparticles confirmed that they were bound by a mixture of (111) facets (white trace) and (100) facets (red trace), corroborated by indexing (e,f) the fast Fourier transform of these regions.*

To give an example of surface energy properties for binary compounds, we plot the Wulff construction for a few Ni-Al systems in Fig. 7, along with single-element Al results for two different FFs. As it is evident from the figure, the surface energies are very much different between the Mishin's Ni-Al[32] potential and NiAlH potential[49], leading to very different Wulff plots. This raises the critical question of which of these FFs should be chosen to perform a Ni-Al MD simulation involving surfaces. In cases like this, without direct experimental (or DFT), a user could decide in terms of which potential reproduces elastic properties better, or which one predict the Wulff plot for the single elements in the binary compound better. The difference in surface energies between DFT and FF implies that dynamic processes can be different between the two modeling techniques, leading to differences in material phenomenon description. Generally, a user should select a FF which has properties similar to DFT, however, if this is not the case, the surface energy prediction in our database can provide insight into why a particular FF behaves different compared to DFT. In addition, if the FF surface energy is close to DFT, then the results should be trustable, and the FF can be used to study relatively large size surfaces, which are difficult to simulate in DFT, and various other surface related phenomenon such as interfaces, stacking faults and grain-boundaries and their temperature dependence.

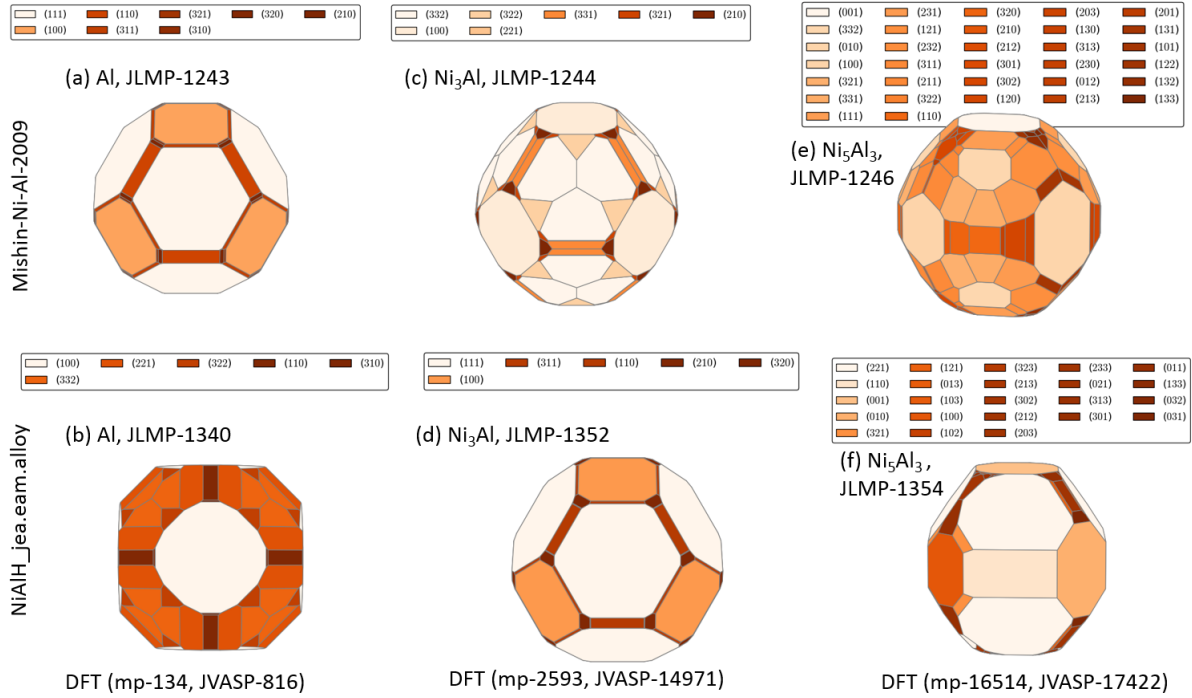

*Fig. 7 Examples of Equilibrium Wulff shape of particles for unary and binary systems with Ni-Al force-fields[26,32]. Wulff-plot for Al, Ni$_3$Al and Ni$_5$Al$_3$ systems are given in a-b, c-d, and e-f respectively. The JLMP- indicate the JARVIS-FF identifier, which can be used to access corresponding webpage, for example,* https://www.ctcms.nist.gov/~knc6/jsmol/JLMP-1243.html *. The materials-project, and JARVIS-DFT identifiers are given by mp-# and JVASP-# from which DFT results can be obtained. These identifiers also lead to corresponding unique webpages such as* https://www.materialsproject.org/materials/mp-134 *,* https://www.ctcms.nist.gov/~knc6/jsmol/JVASP-816 *. The a,c,e results are obtained with FF from Nishin-Ni-Al-2009[25] and the b,d,f results are obtained with FF from NiAlH_jea.eam.alloy[22]. Surface energies are shown by the corresponding colors in the legend for each Wulff-plot.*

In principle, the choice of FF is not the only factor for determining the reliability and reproducibility of MD results. For instance, the simulation set-up parameters and structure-minimization methods can influence the surface and vacancy formation energy values. Hence, we compare two different structure-minimization techniques used during geometric minimization as shown in Table 2. The first method uses LAMMPS box/relax method while the second method uses the refine_relax method. The refine_relax calculation statically calculates the ideal lattice constants and elastic constants at a specified pressure. The underlying algorithm works by having LAMMPS evaluate the pressure for a system as given, and at small positive and negative strains.

The full elastic stiffness tensor, $C_{ij}$, is calculated from the change in pressures with respect to the change in strains. Assuming linear elasticity, the pressure of the unstrained system and the elastic compliance tensor, $S_{ij} = C_{ij}^{-1}$, are used to guess a new box size. This process is then repeated until the lattice constants converge. As evident from the table, the differences between computed values are very small in almost all cases, and definitely much smaller than the differences between various FFs in Fig. 4. We are still investigating sources for these small differences. For instance, the difference could be due to the typical selection of cut-off used during force-field development. Generally, a FF should have reproducible data compared to experiment or DFT, and should be stable to different calculation methods used for calculating properties.

*Table 2. Comparison of refine and box method for FCC Al vacancy and (111) surface energy.*

| Force-field file name | JARVIS-FF | iprPy | JARVIS-FF | iprPy |
|---|---|---|---|---|
| | $V_{Al}$ (eV) | $V_{Al}$ (eV) | $\gamma(111)$ (J/m$^2$) | $\gamma(111)$ (J/m$^2$) |
| Mishin-Ni-Al-2009.eam.alloy[32] | 0.674 | 0.675 | 0.875 | 0.871 |
| Al.set[33] | 0.740 | 0.665 | 0.909 | 0.908 |
| alpb-setfl.eam.alloy[34] | 0.672 | 0.673 | 0.952 | 0.877 |
| Al_wkg_MSMSE_2009.set[35] | 0.662 | 0.663 | 0.88 | 0.876 |
| mg-al-set.eam.alloy[36] | 0.687 | 0.688 | 0.884 | 0.880 |
| Al1.eam.fs[37] | 0.658 | 0.658 | 0.426 | 0.428 |
| Al90Sm10_MendelevM_2014.eam.fs[38] | 0.754 | 0.755 | 1.037 | 0.968 |
| NiAl02.eam.alloy[24] | 1.262 | 1.262 | 1.017 | 0.958 |
| Al-Fe.eam.fs[39] | 0.676 | 0.677 | 0.137 | 0.140 |
| Mishin-Al-Co-2013.eam.alloy[40] | 0.674 | 0.675 | 0.949 | 0.870 |
| NiAl.eam.alloy[41] | 0.707 | 0.708 | 0.593 | 0.601 |
| Al-Mg.eam.fs[27] | 0.658 | 0.658 | 0.426 | 0.428 |
| NiAlH_jea.eam.alloy[26] | 0.524 | 0.525 | 1.018 | 1.000 |
| Farkas_Nb-Ti-Al_1996.eam.alloy[25] | 1.167 | 1.169 | 1.443 | 1.439 |
| Mishin_updated-Al-Co-2013.eam.alloy[40] | 0.674 | 0.675 | 0.875 | 0.871 |
| Mishin_updated-Ni-Al-Co-2013.eam.alloy[40] | 0.674 | 0.675 | 0.875 | 0.871 |
| Experiment[50,51] | 0.68 | 0.68 | 0.855 | 0.855 |

In addition to the classical force-field simulation, our database has straightforward connection to other multi-scale simulation methods. One of the common examples is a phase-field simulation where vacancy-formation energy, surface energy and elastic constant data are required to model material specific evolution of microstructures[52-54]. Such multi-scale simulations will be carried out in future.

**Conclusions:** The JARVIS-FF database comprises the largest collection of consistently calculated elastic, surface and vacancy formation energies data using interatomic potentials to date. We believe that this dataset, the computational framework and the user-friendly webpages will provide a powerful tool in fundamental and application-related studies of materials using classical simulations. Our database allows users to pre-select a force-field prior to a particular classical calculation. We plan to add diffusion, thermal expansion and other interesting properties to our database to evaluate much wider aspects of force-fields in future. Similar tools can be applied to bio/polymer-force-fields as well. We believe integration of classical, quantum and experimental data can enable rapid materials discovery and applications.